# Understanding the Impact of Customer Reviews on Hotel Rating: An Empirical Research

A PREPRINT

**J. Ahmad**
Dept. of Computer Science
*University of Gujrat, Gujrat, Pakistan*
Email: jawad.ahmad@uog.edu.pk

**H. Sami Ullah**
Dept. of Computer Science
*University of Gujrat, Gujrat, Pakistan*
Email: hamzasamiullah@gmail.com

**S. Aslam**
Dept. of Computer Science
*University of Gujrat, Gujrat, Pakistan*
Email: samiaaslam479@gmail.com

## Abstract

The ascent of the Internet has caused numerous adjustments in our lives. The Internet has radically changed the manner in which we carry on with our lives, the manner in which we spend our occasions, how we speak with one another day by day, and how we buy items. The development of the Internet among users has created content on the Internet by sources, for example, web-based life, reviews site, online journals, item fan page and some more. This has a lead on to another method for arranging an occasion or searching for a reasonable hotel to remain. Thus, hotel review sites have turned into a famous stage for visitors to share their experiences, reviews, and suggestions on hotels, which they have visited. In Europe, the hotel business has been a standout amongst the most vital monetary developments of the nation. The essential objective of a hotel is to satisfy the customers, to have the capacity to give a high caliber of administration and give them a vital affair while remaining at the hotel. The motivation behind this examination is to comprehend and recognize the scope of elements, which may add as per the general inclination of customers and in addition through their reviews to decide the measures of customers' desires. Information was gathered from online review sites, for example, Booking.com. Text analytics is utilized to analyze the contents gathered.

## 1. Introduction

Travel planning and booking hotels on the web have turned out to be one of its most vital business employment. With the ascent of the web, the customer's produced reviews, comments, and reports about their visits or tour encounters are playing a growing role as an information source. The customer would commonly post their views towards the hotel and explain encounters on survey sites after their stay at a hotel [1]. However, before booking a hotel on the web, customers will give careful consideration to online reviews of visitors who have experience of staying in a hotel on a social media or any other review stage and will also depict the genuine experience encounter online after they check-in. The online surveys of hotel customers on the web mirror the contrasts between the genuine encounter and the past belief for the customers [2]. Particularly for hotel booking, such customer reviews are pertinent since they are more genuine and nitty-gritty than surveys found in conventionally printed hotel guides etc., they are not one-sided by promoting contemplations as e.g. the hotels' home pages or index depictions and reflects genuine experiences of visitors.

Almost every web travel agency and hotel booking administration these days offer likewise rating for evaluations as well as reviews of hotels, it isn't that simple for hoteliers who need to know what information is spread about their hotels on the web to accumulate the customer produced data. A standard web search engine Google will give a large number of hits for a hotel. However there is a huge number of websites that are providing customer reviews, but these sites are getting data from similar sources, for example, booking.com or tripadvisor.com. In different cases, the links lead us to some broad page from which one can get reviews other than the data and lacking the straightforward route structure of sites. Additionally, the links may point to some individual review keeping it open whether there are different reviews on the site. An extra issue is that the web gives an extensive number of distribution types: other than travel agencies and hotel booking administrations there are various websites, for newsgroups, informal organizations and so forth related to traveling.

Another issue concerns the type of information: travel agencies and hotel booking administrations frequently just distribute scalar ratings, e.g. scores somewhere in the range of 1 and 5. Such scores are not exceptionally accommodating for hotel directors as the numeric values do not give data of what visitors really thought about it whether they are satisfied or not. Likewise, the numeric scores cannot be compared: when a 3-star hotel gets a higher score than a 4- star hotel that does not infer that the one is superior to the next. For hotel managers, the printed customer reviews would be considerably better than the numeric scores since they would be intrigued to comprehend what the customer precisely remarked on and how they thought of it.

One problem for hotel managers is the following updates and new reviews. Hotel booking service and travel agencies gather and distribute customer reviews efficiently, e.g. by approaching their customer reviews and comments by asking them. In this way, new reviews show up much of the time on their front pages still it is hard to follow these by simply using general search. For the traveling user who is getting the reviews on the web for arranging his trips or tours, a large number of these contemplations are not significant, as he will be content with a mentioned depiction of reviews on the hotel's web page. On account of online reviews, researchers have come to end that web-based sentiments are a decent intermediary for informal exchange [3]. In any case, for hoteliers inspired by customer remarks on the web, an administration that consequently and deliberately gathers and condenses the significant data from the web would be worthwhile and maybe much more valuable than the paper forms many hotels use for collecting feedback reports from their visitors.

Sentiment Analysis (additionally called opinion mining) is a sub-field of a text mining and recent research area which can extract emotions or feelings from user-produced content, as hotel reviews into the positive or negative polarity [4], [5]. It is one way to consider the content of surveys [6]. With the polarity and reach of online reviews and comments stages developing quickly, companies are under expanding strength to keep up perfect online popularity [7]. Ongoing years have seen quick development in online discussion groups and reviews sources (e.g.www.tripadvisor.com) where an imperative characteristics of a customer reviews is their feeling or generally speaking sentiment for instance, if the review contains words like 'extraordinary', 'best', 'pleasant', 'great', 'amazing' is most likely a positive remark. While if the review contains words like 'awful', 'poor', 'bad', 'more awful' is presumably a negative comment.

The aim of this paper goes for providing such a service for hotel managers that gather customer reviews for hotels from the web, analyze and classify the textual content of the reviews and present the results in an efficient way.

## 2. Related Work

### 2.1. Data Analytics in Hotel Industry

Data analytics has reformed the cordiality business be it through enhancing by and large customer encounters and expanding consumer loyalty or enhancing business tasks [8]. To use on this potential, information must be put to viable use; information which is the most important to business would be the information which gives bits of knowledge into business activities, obtaining forms and budgetary execution and permit basic decision making progressively exact and breakthrough. The hotel and hospitality industry takes into account millions of travelers. Every customer goes to your hotel with their very own set of desires and needs, which makes it troublesome for hotels to live up to their desires in a generous way. In this way, the hotel business is presently swinging to cutting-edge scientific answers for intimations on approaches to satisfy users and guarantee their customers are satisfied.

### 2.2. Customer Experience and Customer Satisfaction

Customer loyalty is a necessary part of the hotel business [9]. The hotel industry can just thrive and succeed through the capacity to hold customers [1]. Hence, if customers from a hotel don't get benefits up to their standard, they would proceed onward to the following hotel looking for better administrations somewhere else [9]. Satisfaction is considered as the assessment, which customer has encountered with administrations, is comparable to publicized; while another expressed that consumer loyalty alludes to the enthusiastic reaction, an individual has towards an item or administration [10]. Past encounters of explorers will change their desires towards a hotel, and with the moment access to online hotel reviews, it will have a progressively significant effect directing visitors' desires towards a hotel [11]. Client experience can be characterized as a bond and relationship framed between the organization and its client meaning it immediately affects customer loyalty [12]. It is urgent for an organization to manufacture a solid association and leave a decent memory in the psyches of their customers, these recollections will upgrade the encounters of each individual client which will result in satisfied clients while likewise enhancing the dedication of clients [12]. Expressing satisfaction comes just through great, it can fortify the connection between an organization, and a client can make a determination. Organizations are urged to go the additional mile to convey an ordeal like no other and surpass the desires for clients, which may prompt client maintenance.

### 2.3. Text Analytics

The importance of looking into your customer's voice has been settled and advanced by analysts [13]. By analyzing content information, it empowers hotel administrators to better understand the insights of the content and in addition enhance basic decision making. Data Analytics is the change of unstructured content into important information [14]. It centers around extricating the key and significant snippets of data from content information, for example, reviews, discussions, and comments [15]. Through the comprehension of the dialects, content examination can find concealed bits of knowledge of the ”who”, ”where”, and ”when” of every discussion and in addition the ”what” and the ”buzz” of the theme of discussion, ”how” individuals associated with every discussion are feeling, and ”why” is this discussion happening and it is a chance to see each part and to discover importance in every discussion [14]. Data examination is critical to disclosing shrouded bits of knowledge and learning to better understand customers and enhance generally speaking business performance.

# 3. Research Methodology

## 3.1. Identification

”Booking.com” was used to do this exploration. Popular hotels in Europe were chosen. We chose to slither on a survey that influences customer satisfaction. Why? This is to make a better understanding, comprehension of the components, which influences customer satisfaction.

## 3.2. Data Collection

The effective nature of customers on Booking.com demonstrates that it contains a huge collection of reviews and comments. The comments and reviews were gathered from the long stretch of 2015 to 2017, an aggregate time of 2 years with around 500000 reviews.

## 3.3. Cleaning and Analyzing

Text Parsing was used to breaks the terms into different parts, for example, Verb, Noun, and Adjectives. Which terms are used the most depending on frequencies and frequency for its parts were found through text parsing. Its effectiveness can be increased by concentrating on key terms to accelerate the analysis procedure. In the meantime, the text filter is a procedure to recognize terms and separate to refine them as indicated by the implicit lexicon, rehashed terms, stop-words, and so forth. The weight and frequencies can be determinants for the significance of each term and the connections it might have with another. An Interactive Viewer under Text Filter enables us to decide the concept connecting between different terms to distinguish relationship and pattern. For instance, taking the term Room under consideration will enable us to drill down to the key terms, which are related to Room and distinguish hidden patterns and examples. Text Cluster is the way toward gathering comparative terms together. Gathering terms will indicate concealed bits of knowledge, which enables us to recognize the primary topic or thought. Text Topic is to gathering and joins terms into subject collections. Subject collection must be shaped if there are adequate terms, which are corresponded or reliably referenced. Text Topic enables us to exceptionally isolate the subjects dependent on the terms utilized in customer reviews.

## 3.4. Visualization

Graphical visualization was produced through Matplotlib and Seaborn the Python libraries used for data visualizations. Unseen messages can draw out from a graphical representation of data and can give an outline analysis of the term connecting.

# 4. Descriptive Analysis

This dataset contains 500000 customer reviews and scoring of 1493 luxury hotels across Europe. In the meantime, the geographical area of hotels is also provided for further analysis.

## 4.1. Data Content

The dataset file contains 17 fields. The description of each field is as below:
Hotel Address: Address of the hotel.
Review Date: Date when the reviewer posted the corresponding review.
Averages Score: Average Score of the hotel, calculated based on the latest comment in the last year.
Hotel Name: Name of Hotel
Reviewer Nationality: Nationality of Reviewer
Negative Review: Negative Review the reviewer gave to the hotel. If the reviewer does not give the negative review, then it should be: 'No Negative'
Review Total Negative Word Counts: Total number of words in the negative review.
Positive Review: Positive Review the reviewer gave to the hotel. If the reviewer does not give the negative review, then it should be: 'No Positive'

Review Total Positive Word Counts: Total number of words in the positive review.
Reviewer Score: Score the reviewer has given to the hotel, based on his/her experience
Total Number of Reviews Reviewer Has Given: Number of Reviews the reviewers has given in the past.
Total Number of Reviews: Total number of valid reviews the hotel has.
Tags: Tags reviewer gave the hotel.
Days since review: Duration between the review date and scrape date.
Additional Number of Scoring: There are also some guests who just made a scoring on the service rather than a review. This number indicates how many valid scores without review in there.
lat: Latitude of the hotel
lng: longitude of the hotel

In the first step, we cleaned the data observing that there were 526 duplicate records so we removed them from the data. We also observed that there were 3268 missing values in lat, lng features. The data of 17 hotels contained the missing values so we didn't want to lose the data. Therefore instead of removing these missing values we first tried to find those hotels with missing values in remaining data if found Then we can fill

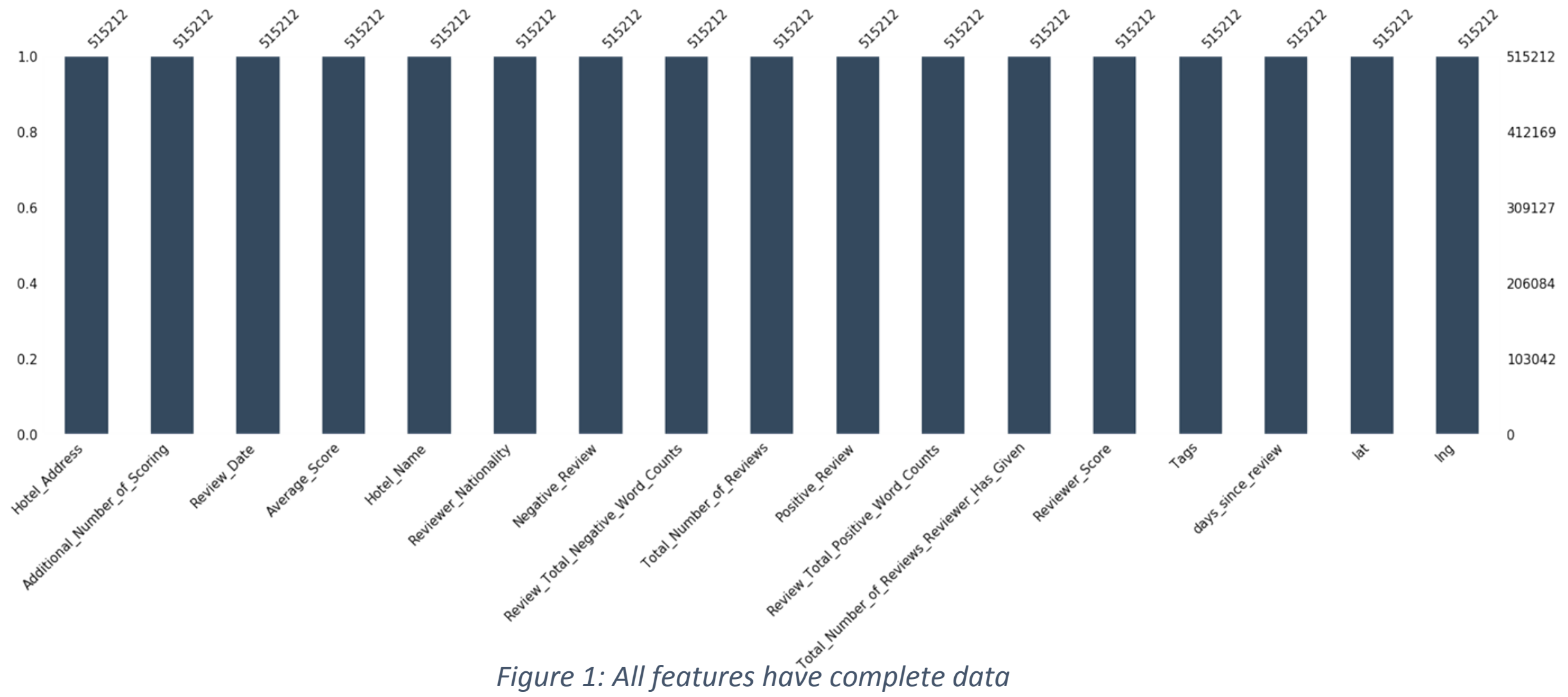

*Figure 1: All features have complete data*

those missing values with them. But those hotels couldn't be found so we manually fill the missing values by getting lat lng from this source [16]. We can see that no feature contain any missing value in Figure 1:

In the above Figure, 1 column names are given below and the top values show the number of values in each column left values are of a scale and right values are the parts of total record. The score of hotel matters a lot in hotel ranking the range of average score is from 5.2 to 9.8 and how hotels have the average score is effectively shown in the figure below:

Our hotel data belong to some prominent countries in Europe. Europe is well established in the hotel industry and business. A huge number of people go to Europe for the tour each year. They like to stay in hotels there. All the reviews data of hotels belong to 6 countries Italy, Austria, Netherland, France, Spain, and the UK. What is the hotels' ratio in these countries is shown in Figure 2 given below:

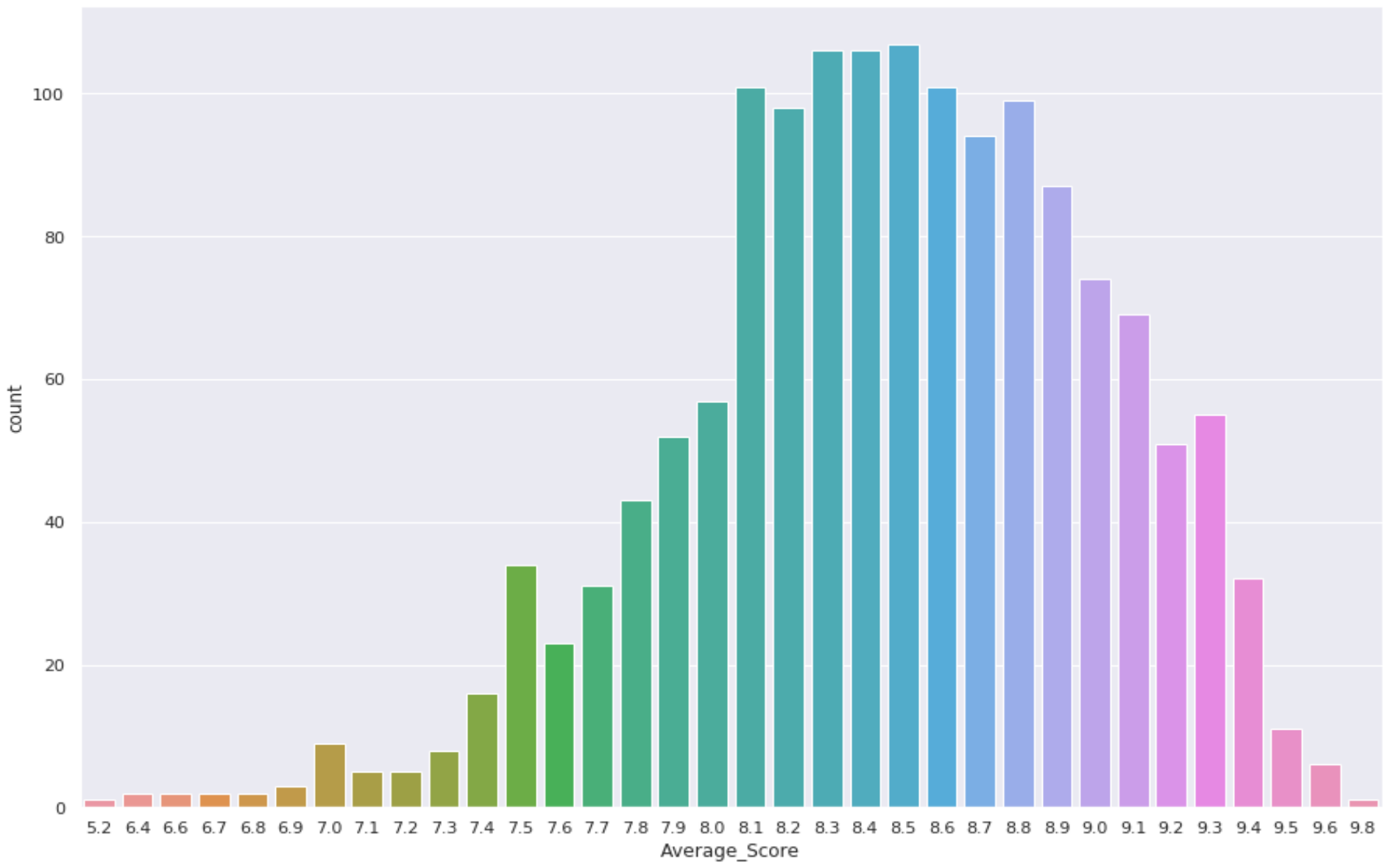

*Figure 2: Ratio of hotels on average score*

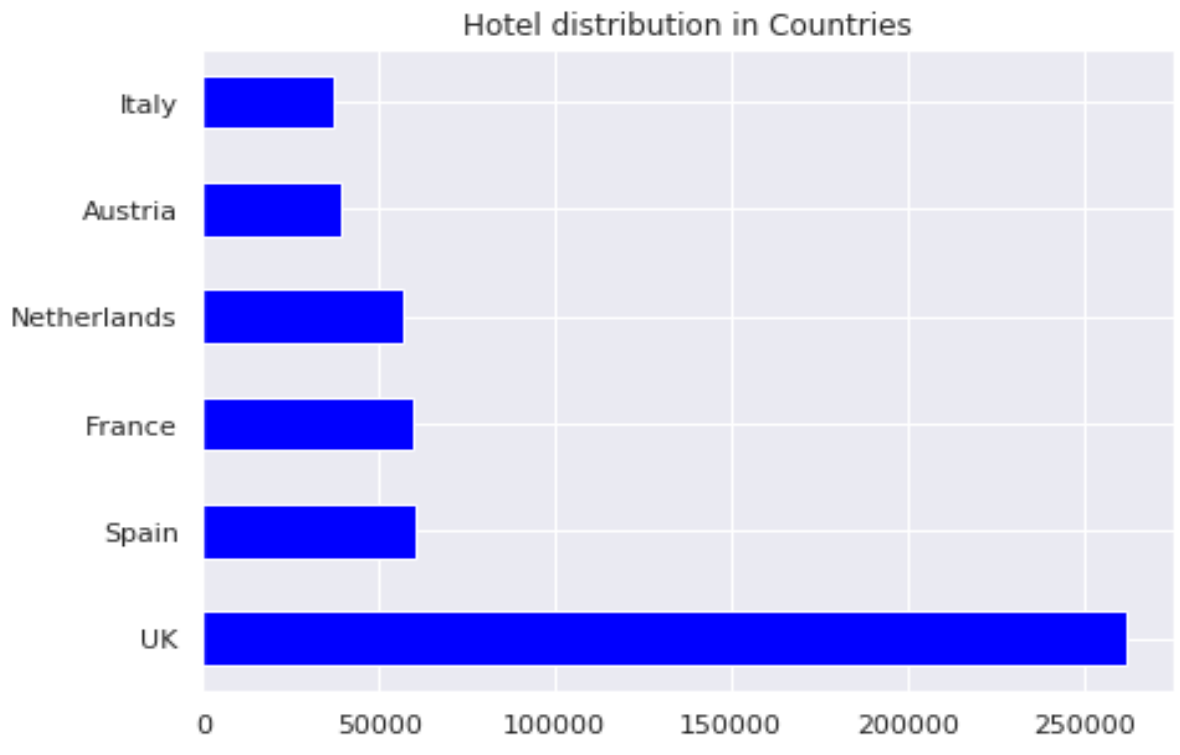

Now we talk about the data collection from which sources are actually data providers. We downloaded this data set from kaggle.com. But in the data who are the people giving the reviews. We just need to know about them. They can belong to any country, nation, state or any culture. To have look at the nationality of the reviewers a Figure 3 visualize their nationalities.

For the quick analysis to know the performance of the hotels or hotel ratings. We have a very useful feature in the dataset that is the average score. If we want to look at the best and worst hotels in Europe from the given dataset, we can visualize the hotels according to their average score. We also visualize the top 10 hotels in both cases worst and best, based on their average score. The average score of the hotels is a quick way to observe the hotel status. Although it should not be considered as a very authentic source to judge the hotel ranking or its status. There can be a lot of factors involved in this decision.

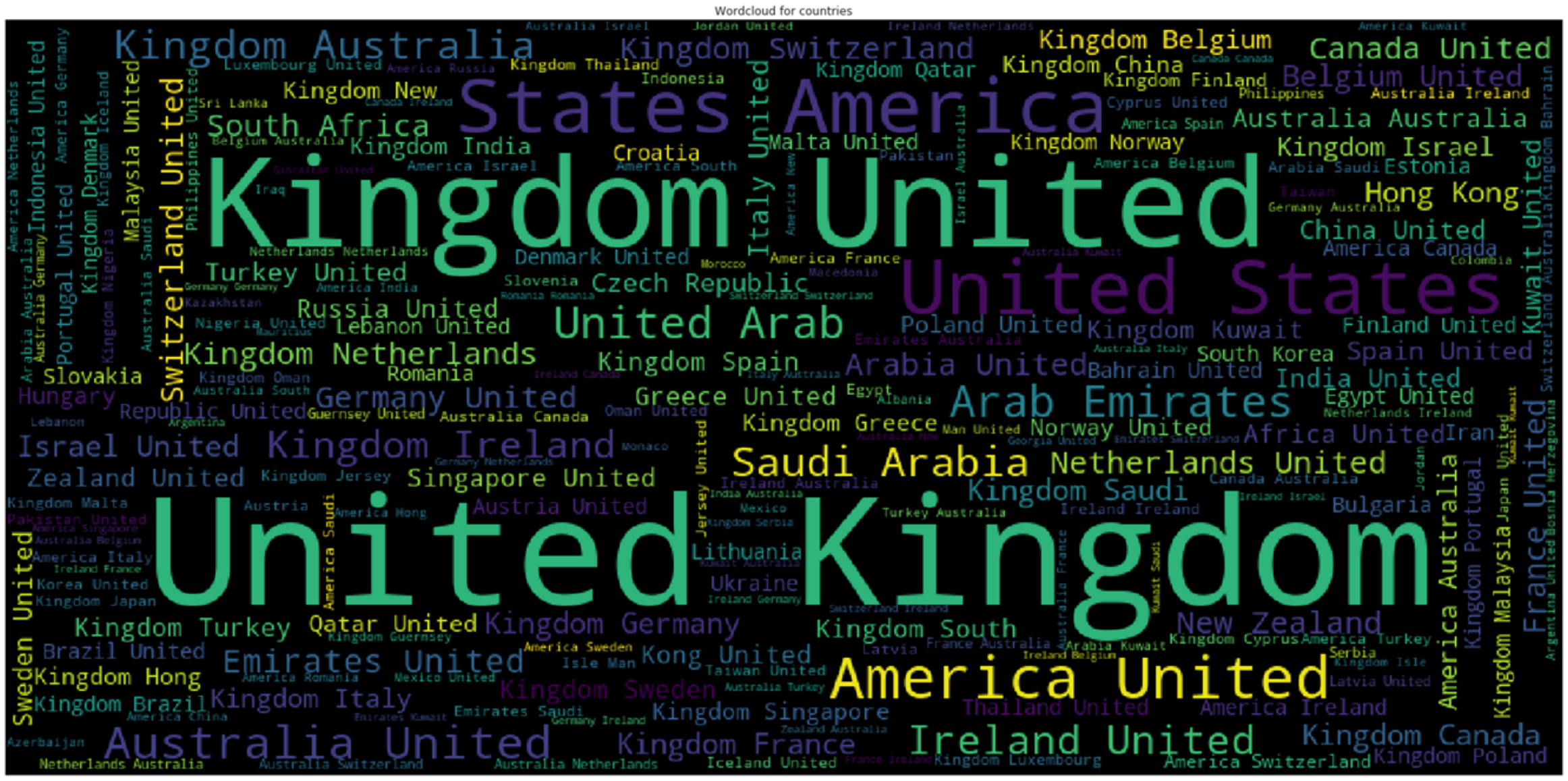

*Figure 3: Reviewer nationalities word cloud*

Top 10 best hostels with most of the average score and top 10 worst hotels with least of the average score are represented in the below bar charts.

Now, what about the time and date of the review submission. How many reviews were given in each year of 2015, 2016 and 2017 and each month of all three years? It is very compulsory to aware of the reviews time. It helps us in the process of data collection for further experiments. We would know about the more important dates on which the reviews are given. Here is the representation of a number of reviews in terms of time.

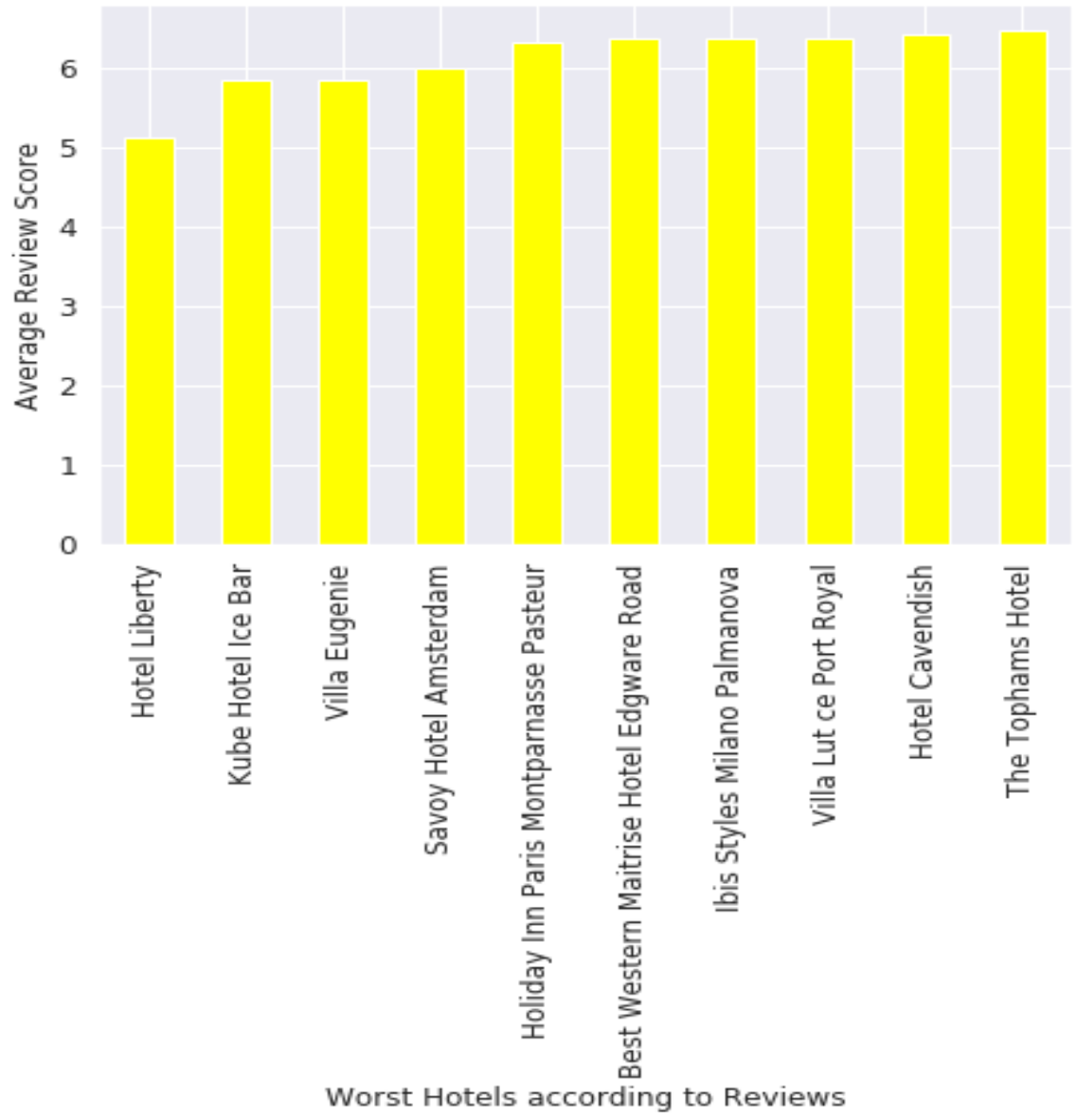

*Figure 4: Top 10 worst hotels according to average score*

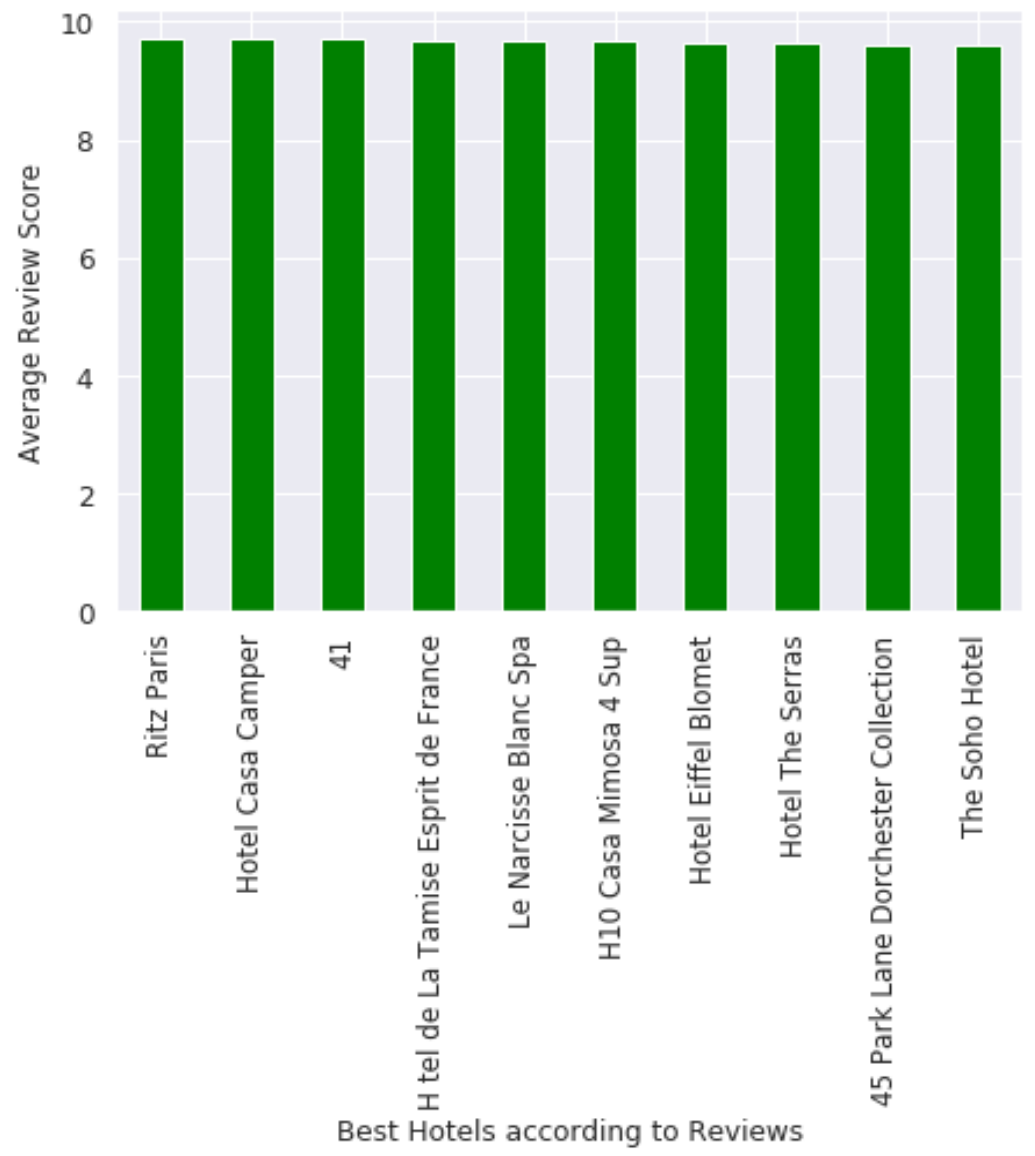

*Figure 5: Top 10 best hotels according to average score*

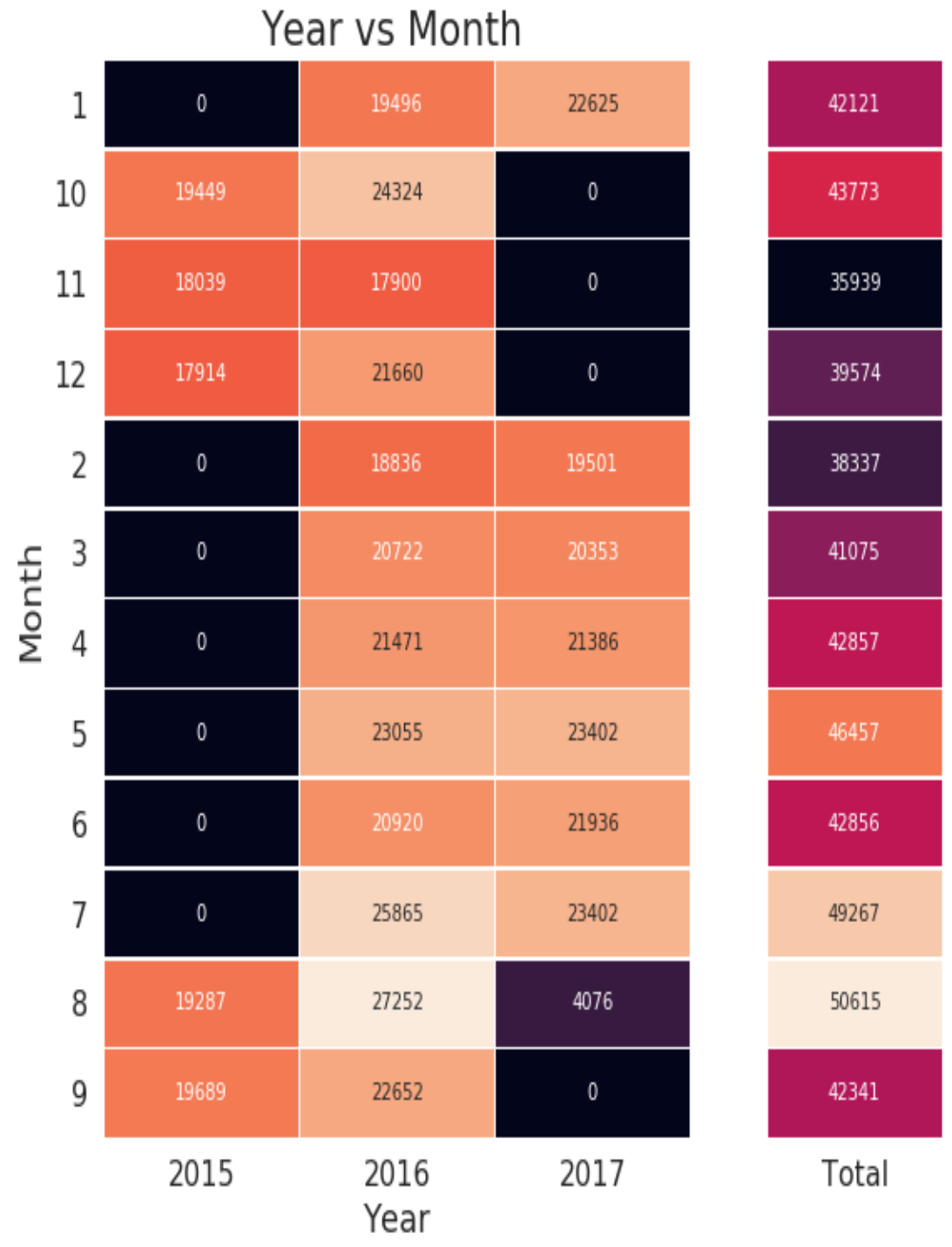

*Figure 6: Year-wise and month-wise calculations*

A number of reviews also have a great factor in any hotel to be considered as a well-known business industry. Although these are the positive or negative first thing to observe is that this is what about people like or want to say something. This is the point to be noted. In this way, that hotel gets into the public discussions. The figure below shows the top 10 reviewed hotels.

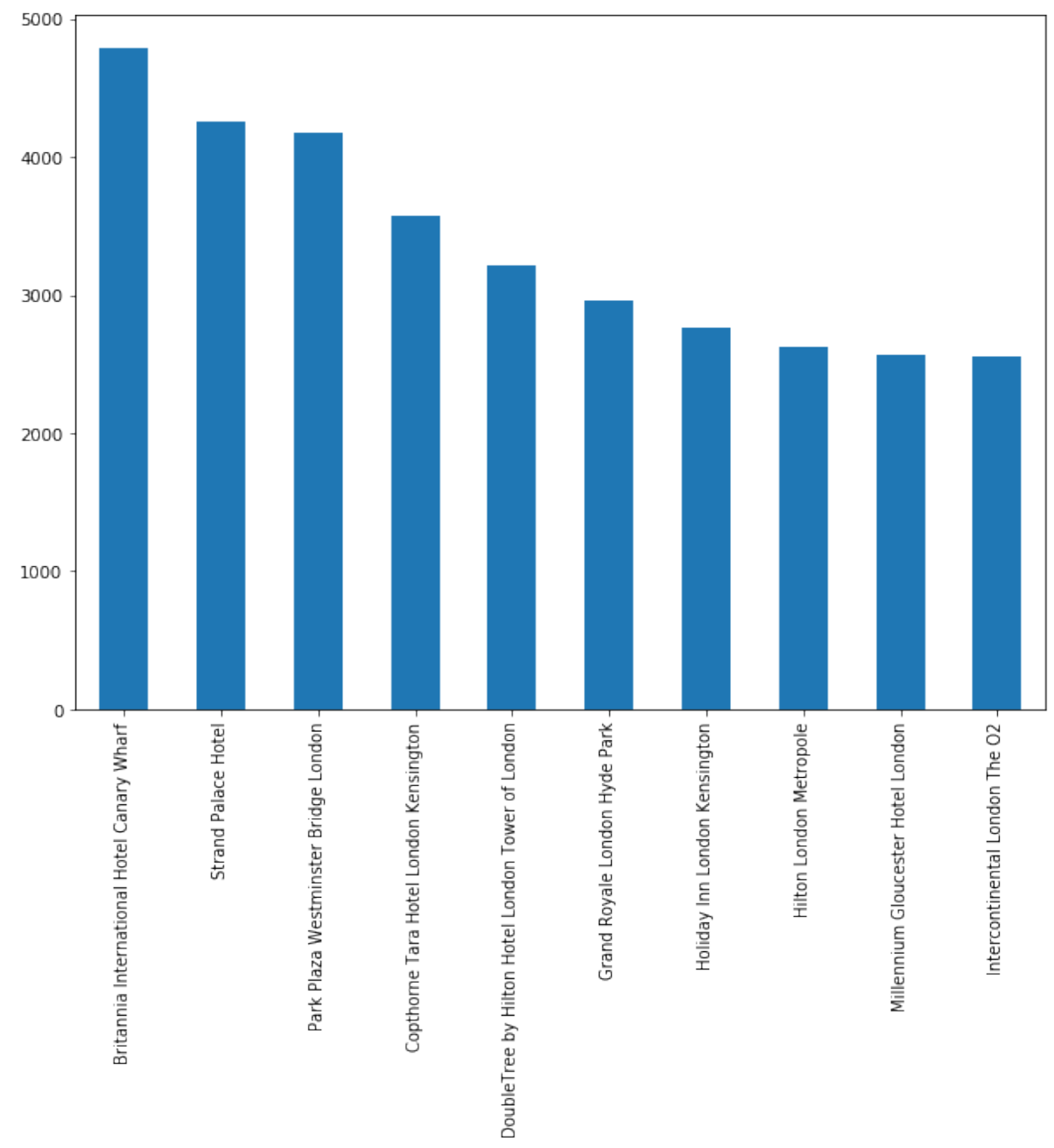

*Figure 7: 10 Reviewed Hotels*

We have extracted some important words from the reviews. Words on which most of the viewers have great attention are very useful for prediction purposes and the business trend of that particular hotel. Some people mentioned that important word and some don't. We need to visualize important words according to the average score.

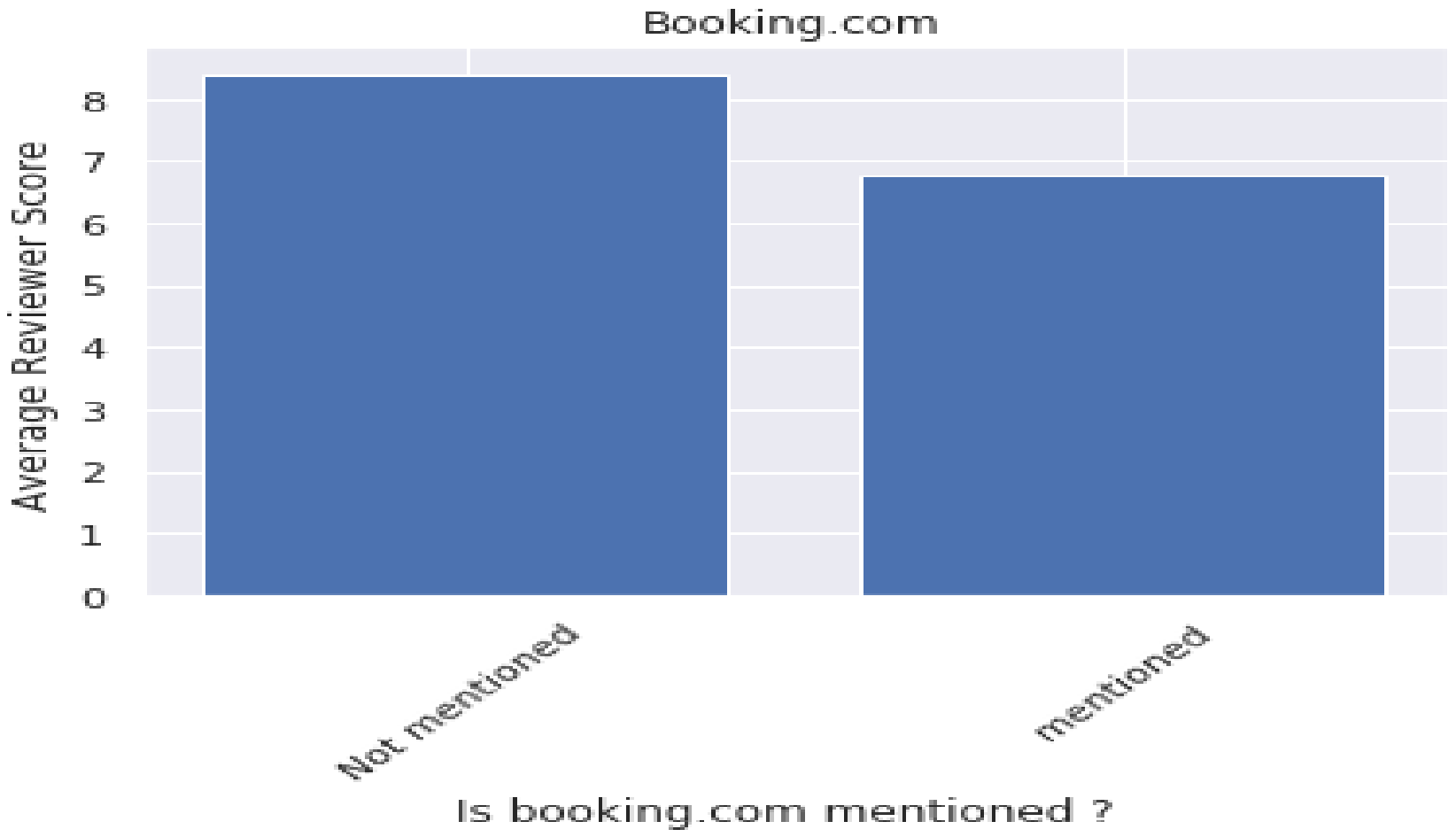

*Figure 8: Booking Reviews*

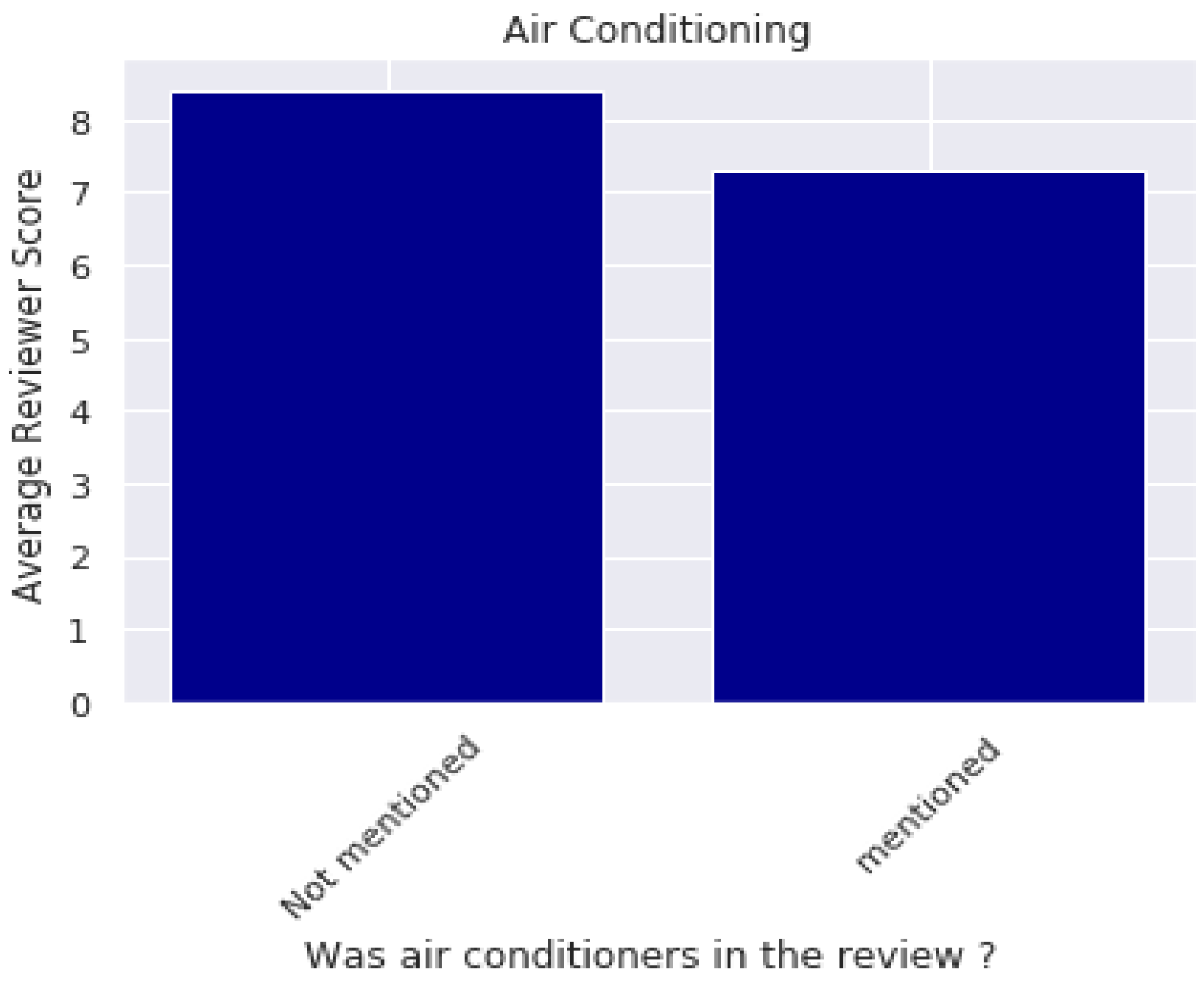

*Figure 9: Air Conditioning Review*

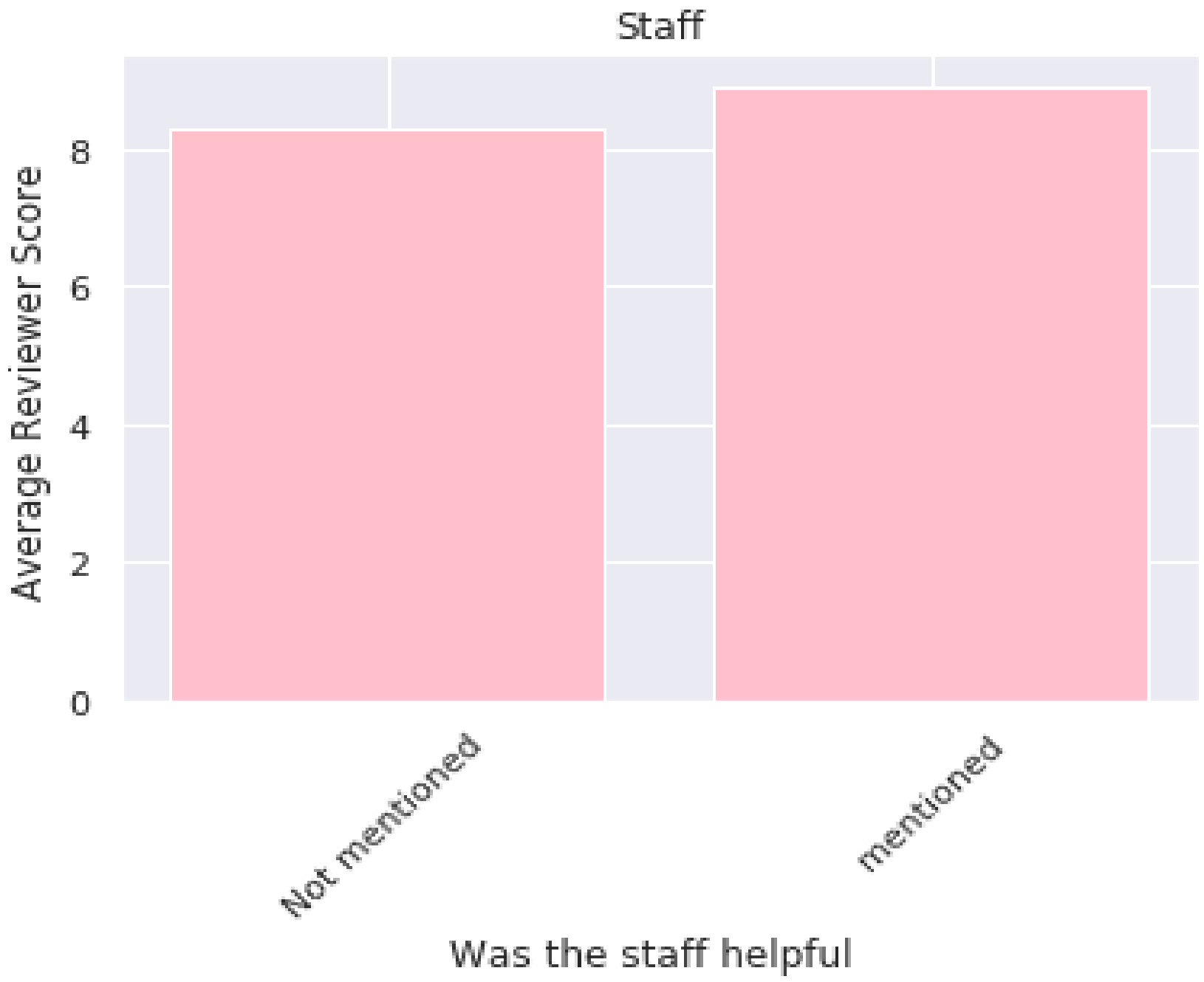

*Figure 10: Staff Review*

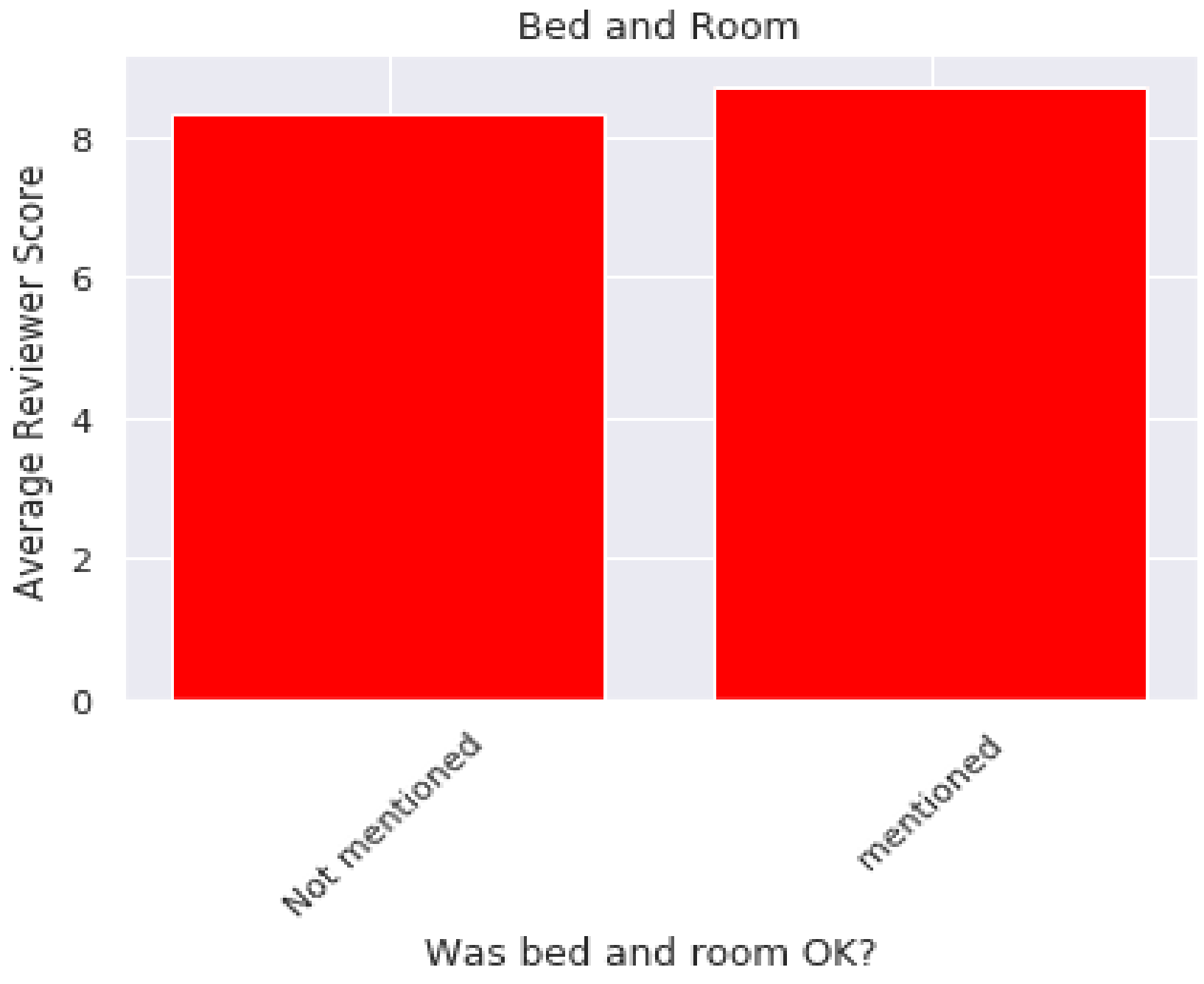

*Figure 11: Bed and Room Review*

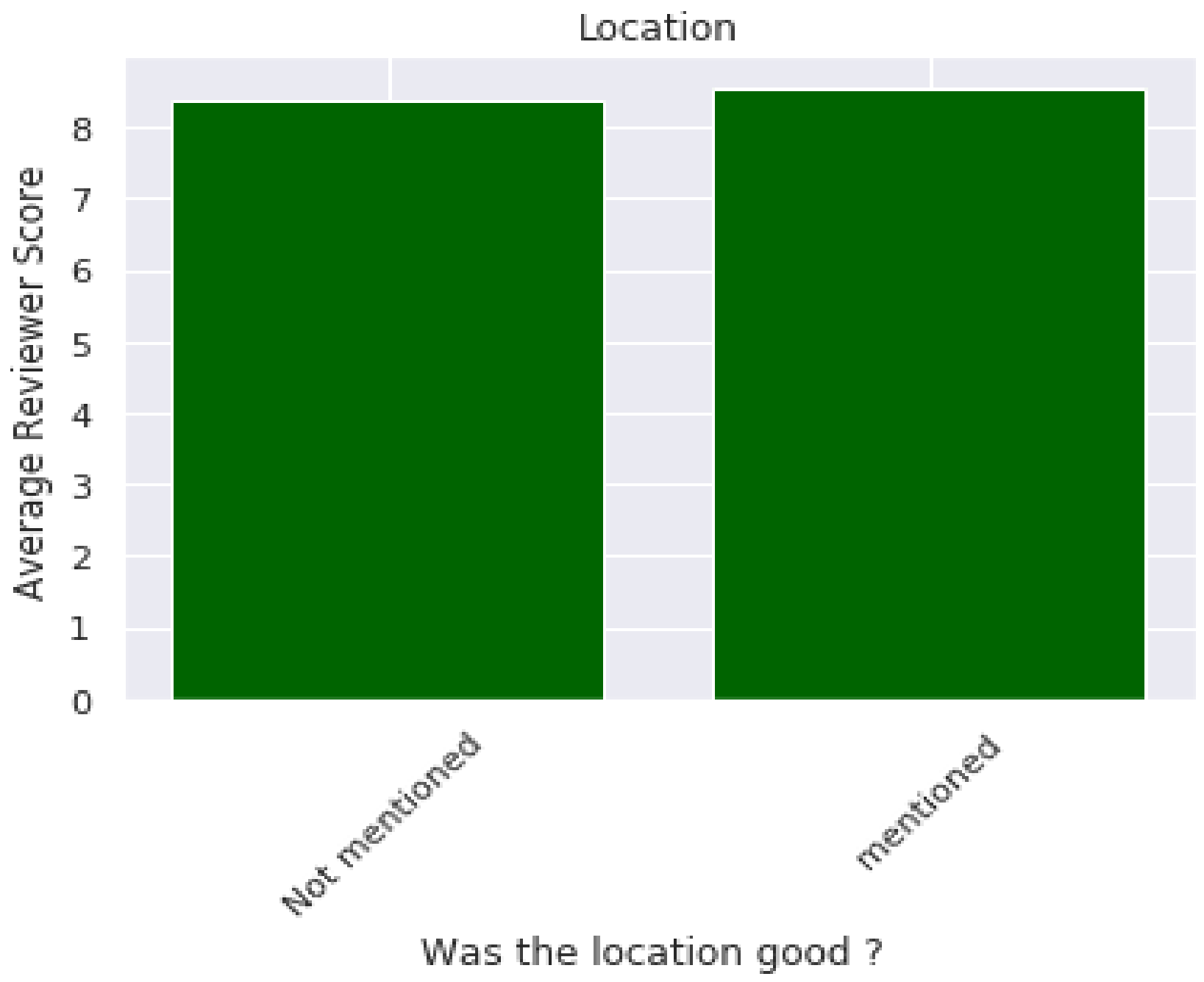

*Figure 12: Location Review*

## 5. Limitations and Future Research

The strategy was reinforced by considering the popular hotels in Europe; in any case, the aftereffects of this examination must be seen under a couple of impediments. Right off the bat, the information has just been gathered for hotels and for a long time, which restricts the generalizability of the outcomes to different hotels in the travel business. Future investigations are prescribed to test a couple of more hotels or maybe a similar chain of hotels in an alternate country.

Furthermore, the examination has just shown the consequences of topic connecting in a review, which confines one on knowing the Inverse Document Frequency (IDF). Future investigations are prescribed to mimic the investigation of term loads and the measurement decrease with SVD to have the load an incentive between each term where this will decide the significance of each term points.